\documentclass[twocolumn,amsmath,amssymb,prl,superscriptaddress,aps,showpacs,10pt]{revtex4-1}

\usepackage{amsfonts}
\usepackage{graphicx}

\newcommand{\kb}[1]{\textrm{\tiny{#1}}}
\newcommand{\kc}[1]{\mathbf{#1}}
\newcommand{\kd}[1]{\mathbf{#1}}

\begin{document}
 
\title{Ultra-long-range giant dipole molecules in crossed electric and magnetic fields}

\date{\today}
\author{Markus Kurz}
\affiliation{Zentrum f\"ur optische Quantentechnologien, Luruper Chaussee 149, 22761 Hamburg, Germany}
\author{Michael Mayle}
\altaffiliation[Present address: ]
{JILA, University of Colorado and National Institute of Standards and Technology, Boulder, Colorado 80309-0440, USA}
\affiliation{Zentrum f\"ur optische Quantentechnologien, Luruper Chaussee 149, 22761 Hamburg, Germany}
\author{Peter Schmelcher}
\affiliation{Zentrum f\"ur optische Quantentechnologien, Luruper Chaussee 149, 22761 Hamburg, Germany}
\begin{abstract}
We show the existence of ultra-long-range giant dipole molecules formed by a neutral alkali ground state atom that is bound to the decentered electronic wave function of a giant dipole atom. The adiabatic potential surfaces emerging from the interaction of the ground state atom with the giant dipole electron posses a rich topology depending on the degree of electronic excitation. Binding energies and the vibrational motion in the energetically lowest surfaces are analyzed by means of perturbation theory and exact diagonalization techniques. The resulting molecules are truly giant with internuclear distances up to several $\mu m$. Finally, we demonstrate the existence of intersection manifolds of excited electronic states that potentially lead to a vibrational decay of the ground state atom dynamics.
\end{abstract}
\maketitle
\section{Introduction}

Since the discovery of Bose-Einstein condensates \cite{Anderson95,Davis95} the knowledge of ultracold atomic and molecular systems has grown with a breathtaking speed. Experimentally, one can extensively control the external motion of the atoms by designing and switching between almost arbitrarily shaped traps \cite{Pethick08,Grimm00,Folman02} and the strength of the interaction among the atoms can be tuned by magnetic or optical Feshbach resonances \cite{Koehler06,Bloch08,Chin10}. This has not only lead to a plethora of novel observations on the many-body behavior of ultracold atomic ensembles \cite{Bloch08} but also raised new ultracold few-body physics such as weakly bound diatomic molecules \cite{Jochim03} or Efimov states of timers \cite{Knoop09} in the universal regime close to the dissociation threshold. A particularly striking new species are the weakly bound ultra-long-range diatomic molecules composed of a ground state and a Rydberg atom whose existence has been predicted theoretically a decade ago \cite{Greene00} and which have been discovered experimentally only recently \cite{Bendkowsky09}. The molecular Born-Oppenheimer potential energy curves, which are responsible for the atomic binding, show for these species a very unusual oscillatory behavior with many local minima. The latter can be understood intuitively and modeled correspondingly as the interaction of a neutral ground state atom with the Rydberg electron of the second atom. The equilibrium distance for these molecular states is of the order of the size of the Rydberg atom and the vibrational binding energies are in the MHz to GHz regime for principal quantum numbers $n \approx 30-40$. Recently, the impact of magnetic fields on these ultra-long-range molecules \cite{Lesanovsky07} and the formation of Rydberg trimers and excited dimers by internal quantum reflection \cite{Bendkowsky10} has been explored. Moreover, it it has been shown how the electric field of a Rydberg atom electron can bind a polar molecule to form a giant ultra-long-range stable polyatomic molecule \cite{Rittenhouse10,Rittenhouse11} and how Rydberg macrodimers are formed via long-range interaction between two atoms excited into high-$n$ Rydberg states \cite{cote11}.

An exotic species of Rydberg atoms in crossed electric and magnetic fields are the so-called giant dipole states (GDS), which have been explored theoretically \cite{Baye92,Dzyaloshinskii92,Schmelcher93,Dippel94} and experimentally \cite{Fauth87,Raithel93} firstly in the 1990s. Opposite to the usual Rydberg states, the GDS are of decentered character and possess a huge electric dipole moment. More precisely, in ref.\cite{Dippel94} it was shown that the total potential of the electronic motion possesses a gauge invariant term which leads to an outer potential well that supports weakly bound decentered states. The mathematical origin of these effects is the non-separability of the center of mass and electronic motion in the presence of the external fields \cite{Avron78,Johnson83}: translation symmetry and conservation of the total momentum in field-free space is replaced by a phase space translation symmetry and the conservation of the total pseudo momentum. Applications to the Positronium atom have demonstrated that metastable matter-antimatter states with a lifetime of many years can be formed \cite{Ackermann98}. More recently, giant dipole resonances of multiply excited atoms in crossed fields have been shown to exist and the corresponding electronic configurations as well as their stability have been analyzed \cite{Schmelcher01,Zoellner05}.

In the present work we combine the concepts of atomic giant dipole states and field-free ultra-long-range diatomic molecules: We show the existence of ultra-long-range giant dipole molecules emanating from giant dipole states. They exist in a variety of different configurations with simple to complex three-dimensional potential energy surfaces such as Gaussian, elliptical or toroidal wells. For higher excited states the potential energy surfaces of energetically neighboring states come close in configuration space and form higher-dimensional seams of avoided crossings that could lead to rapid decay processes of vibrational wave packets.

\section{Model Hamiltonian}
We consider a highly excited hydrogen atom interacting with a ground state neutral perturber atom (we will focus on the $^{87}$Rb atom here) in crossed static homogeneous electric $\kd{E}$ and magnetic $\kd{B}$ fields. The corresponding Hamiltonian reads 
\begin{eqnarray}
H = \frac{\kc{p}^{2}_{\kb{n}}}{2 m_{\kb{n}}} + H_{\kb{GD}} + V_{\kb{n,GD}}, \label{model1}
\end{eqnarray}
where `n' labels the neutral perturber and where the first term is the kinetic energy of the perturber atom followed by the giant dipole Hamiltonian  of the hydrogen atom and the interaction term between the GDS and the neutral perturber. The hydrogen atom in crossed external fields has been discussed in detail in ref. \cite{Dippel94}. There it was shown that the giant dipole Hamiltonian can be transformed into an effective single particle problem in a magnetic field in the presence of a generalized potential $V(\kd{r};\{ \kd{E},\kd{B} \})=\frac{1}{2M}(\kc{K}-e {\kc{B}} \times {\kc{r}})^2 - \frac{e^2}{r}$ which parametrically depends on the external fields and the pseudo momentum ${\kc{K}}({\kc{E}},{\kc{B}})$ and contains both the motional and external electric field Stark terms,
\begin{eqnarray}
H_{\kb{GD}}=\frac{1}{2\mu}(\kc{p}-\frac{q}{2}\kd{B}\times \kd{r})^2 + V(\kd{r};\{ \kd{E},\kd{B} \}),
\end{eqnarray}
with $\mu=\frac{m_{\kb{e}}m_{\kb{p}}}{m_{\kb{e}}+m_{\kb{p}}},\ \ \ q=\left(\frac{m_{\kb{e}}-m_{\kb{p}}}{m_{\kb{e}}+m_{\kb{p}}}\right)e$ where $e,m_{\kb{e}},m_{\kb{p}},M$ are the electron charge and mass and the proton and atomic mass, respectively. ${\kc{r}},{\kc{p}}$ represent the coordinate and canonical momentum of the Rydberg electron. For sufficiently strong electric fields the potential $V$ exhibits an outer well containing many bound GDS which are decentered from the proton at distances of about $10^5a_0$, $a_0$ being the Bohr radius. The latter leads to a huge electric dipole moment typically of the order of many ten thousand Debye for strong electric and magnetic laboratory field strengths. Expanding $V(\kd{r};\{ \kd{E},\kd{B} \})$ up to second order around the minimum of the outer well and performing a corresponding gauge centering \cite{Schmelcher93,Dippel94} we arrive at the second order giant dipole Hamiltonian which represents a charged (effective) particle in a magnetic field and an anisotropic 3D harmonic potential,
\begin{eqnarray}
H_{\kb{GD}}=\frac{1}{2\mu}(\kc{p}-\frac{q}{2}\kd{B}\times \kd{r})^2 + \frac{\mu}{2}\omega^2_{\kb{x}}x^2 + \frac{\mu}{2}\omega^2_{\kb{y}}y^2 +\frac{\mu}{2}\omega^2_{\kb{z}}z^2,
\end{eqnarray}
where the frequencies $\omega_{x}=\sqrt{\frac{2}{\mu}(\frac{B^2}{2M}+\frac{1}{x^{3}_{0}})}$, $\omega_{y}=\sqrt{\frac{1}{\mu}(\frac{B^2}{2M}-\frac{1}{x^{3}_{0}})}$ and $\omega_{z}=1/\sqrt{|\mu x^{3}_{0}|}$ characterize the anisotropy of the outer well. In this representation, ${\kc{r}},{\kc{p}}$ denote the electronic variables with respect to the outer minimum $(x_0,0,0)$ with $x_0 \approx - |\kd{K}|/B$. Our working Hamiltonian therefore reads
\begin{eqnarray}
H&=&\frac{\kc{p}^{2}_{\kb{n}}}{2m_{\kb{n}}}+\frac{1}{2\mu}(\kc{p}-\frac{q}{2}\kd{B}\times \kd{r})^2 \nonumber \\ 
&\ & + \frac{\mu}{2}\omega^2_{\kb{x}}x^2 + \frac{\mu}{2}\omega^2_{\kb{y}}y^2 +\frac{\mu}{2}\omega^2_{\kb{z}}z^2+V_{\kb{n,GD}}(\kd{r},\kd{r}_{\kb{n}}), \label{model}
\end{eqnarray}
where $V_{\kb{n,GD}}$ again represents the interaction of the neutral perturber atom with the giant dipole electronic Rydberg state. For deeply bound states in the outer well the electron possesses a low kinetic energy and it is legitimate to describe the interaction with the neutral perturber by a Fermi-type pseudopotential \cite{Greene00,Omount}, namely, a $s$-wave contact potential 
\begin{eqnarray}
V_{\kb{n,GD}}(\kd{r},\kd{r}_{\kb{n}})=2\pi A_{\kb{T}}[k(\kd{r})]\delta^{(3)}(\kd{r}-\kd{r}_{\kb{n}}). \label{model3}
\end{eqnarray}
Here $A_{\kb{T}}(k)$ is the energy-dependent triplet $s$-wave scattering length for electron collisions with the ground state Rb atom.  ${\kc{r}}_{\kb{n}}$ denotes the position of the neutral perturber with respect to the minimum of the outer well. The electron wave number $k$ is provided by the kinetic energy of the Rydberg electron when it collides with the neutral perturber \cite{Greene00}.
\section{Methodology}
In order to solve the eigenvalue problem associated with Hamiltonian (\ref{model}) we adopt an adiabatic ansatz for the neutral ground state and the giant dipole atom. We write the total wave function as $\Psi(\kd{r},\kd{r}_{\kb{n}})=\phi(\kd{r}_{\kb{n}})\psi(\kd{r};\kd{r}_{\kb{n}})$ and yield
\begin{eqnarray}
[H_{\kb{GD}}+V_{\kb{n,GD}}(\kd{r},\kd{r}_{\kb{n}})]\psi_i (\kd{r};\kd{r}_{\kb{n}})&=&\epsilon_i
(\kd{r}_{\kb{n}})\psi_i(\kd{r};\kd{r}_{\kb{n}}), \label{calc1}\\ 
 \left[ \frac{\kc{p}^{2}_{\kb{n}}}{2 m_{\kb{n}}}+\epsilon_i(\kd{r}_{\kb{n}}) \right] \phi_k^i(\kd{r}_{\kb{n}})&=&E_k^i\phi_k^i(\kd{r}_{\kb{n}}),\label{calc2}
\end{eqnarray}
where $\psi$ describes the electronic wave function of the decentered GDS in the presence of the neutral perturber for a given position $\kd{r}_{\kb{n}}$ and $\phi$ determines the vibrational state of the neutral perturber. To calculate the potential energy surfaces (PES) $\epsilon_i({\kd{r}}_{\kb{n}})$ we expand $\psi_i(\kd{r};\kd{r}_{\kb{n}})$ in the eigenbasis of $H_{\kb{GD}}$, i.e., $\psi_i(\kd{r};\kd{r}_{\kb{n}})=\sum_{j} C_{j}^i(\kd{r}_{\kb{n}})\chi_{j}(\kd{r})$ with $H_{\kb{GD}}\chi_{j}(\kd{r})=\varepsilon_{j}\chi_{j}(\kd{r})$, and solve the corresponding eigenvalue problem associated with eq.\ (\ref{calc1}) using standard numerical techniques for the diagonalization of Hermitian matrices. In \cite{Dippel94} it was shown that the eigenenergies and eigenfunctions of $H_{\kb{GD}}$ are determined by the three quantum numbers $n_{-},\ n_{+},\ n_{\kb{z}} = 0,1,2,... [\chi_{j}(\kd{r})\equiv \chi_{n_{-}n_{+}n_{\kb{z}}}(\kd{r})]$, where $\epsilon_{n_{-},n_{+},n_{\kb{z}}}=\hbar\omega_{-}(n_{-}+\frac{1}{2})+\hbar\omega_{+}(n_{+}+\frac{1}{2})+\hbar\omega_{\kb{z}}(n_{\kb{z}}+\frac{1}{2})$ with $\omega_{\pm}=\frac{1}{\sqrt{2}}[\omega^{2}_{\kb{x}}+\omega^{2}_{\kb{y}}+\omega^{2}_{\kb{c}} \pm \text{sgn}(\omega^{2}_{\kb{x}}-\omega^{2}_{\kb{y}})\sqrt{(\omega^{2}_{\kb{x}}+\omega^{2}_{\kb{y}}+\omega^{2}_{\kb{c}})^2-4\omega^{2}_{\kb{x}}\omega^{2}_{\kb{y}}}]^{1/2}$ and $\omega_{\kb{c}}=-qB/\mu$. To ensure convergence for our numerical approach we vary the number of orbitals associated with the quantum numbers $n_{-},\ n_{+},\ n_{\kb{z}}$ independently finally achieving a relative accuracy of $10^{-5}$ for the energy. To do so for the energetically lowest 15 excitations a basis set of approximately $1500$ states is needed. In addition to the numerically exact treatment we determine the PES in first order perturbation theory, leading to
\begin{eqnarray}
\epsilon_{j}^{\kb{pt}}(\kd{r}_{\kb{n}})=\varepsilon_{j}+2\pi A_{\kb{T}}[k(\kd{r}_{\kb{n}})]|\chi_{j}(\kd{r}_{\kb{n}})|^{2}.
\end{eqnarray}
From eqs.\ (\ref{model})-(\ref{calc2}) we can already deduce some symmetry properties of the states $\Psi,\ \phi$ and the energies $\epsilon (\kd{r})$. If $P_{\kd{r},\kd{r}_{\kb{n}}}$ denotes the parity operator that transforms $(\kd{r},\kd{r}_{\kb{n}})\rightarrow (-\kd{r},-\kd{r}_{\kb{n}})$ we have
$[H,P_{\kd{r},\kd{r}_{\kb{n}}}]=[V_{\kb{n,GD}}(\kd{r},\kd{r}_{\kb{n}}),P_{\kd{r},\kd{r}_{\kb{n}}}]=0$. This means that the states $\Psi,\ \psi$ and $\phi$ are parity (anti)symmetric and the PES are symmetric, i.e., $\epsilon(\pm \kd{r}_{\kb{n}})=\epsilon(\kd{r}_{\kb{n}})$.

Throughout this work we use the exemplary field configuration $E=50$ V/cm and $B=2.35$ T.     This gives giant dipole level spacings of $\omega_{-}=223\ \text{MHz},\ \omega_{+}=413\ \text{GHz}\ \text{and}\ \omega_{z}=1.35\ \text{GHz}$. For such fields, it is justified to neglect all interaction terms due to the induced dipole of the Rb atom. 
\section{Molecular ground state potential surfaces}
\begin{figure}
\includegraphics[width=\columnwidth]{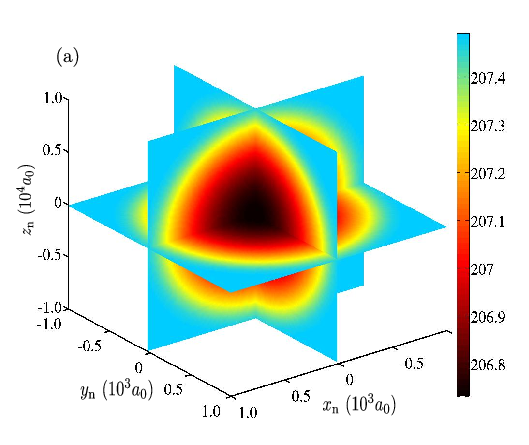}
\includegraphics[width=0.975\columnwidth]{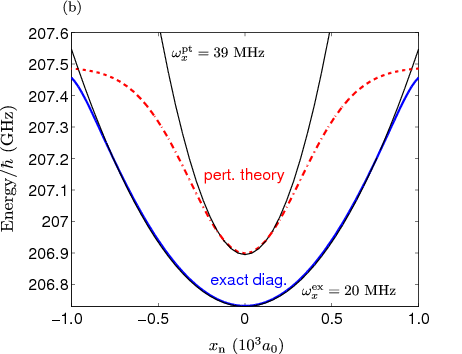}
\caption{(a) Ground state potential energy surfaces calculated via exact diagonalization. The employed basis set consists of $N_{-}=30,\ N_{+}=1,\ N_{\kb{z}}=50$ states. The energy scale is given in GHz. (b) Comparison between perturbative (dashed-dotted line) and exact (solid line) ground state potential for $y_{\kb{n}}=z_{\kb{n}}=0$. In addition, the harmonic fitting curves (thin lines) together with the corresponding trap frequencies $\omega^{\kb{pt}}_{x}$ and $\omega^{\kb{ex}}_{x}$ are shown. (color online)}\label{fig.1}
\end{figure}

In first order perturbation theory the molecular giant dipole ground state PES is given by 
\begin{eqnarray}
\epsilon^{\kb{pt}}_{000}(\kd{r}_{\kb{n}})=\varepsilon_{000}+2\pi A_{\kb{T}}[k(\kd{r}_{\kb{n}})]C_{0}e^{-ax^{2}_{\kb{n}}-by^{2}_{\kb{n}}-cz^{2}_{\kb{n}}},\label{ptexc1}
\end{eqnarray}
where the constants $C_0,a,b,c>0$ are given in \cite{Dippel94}. This potential represents a single well 3D potential and since $A_{\kb{T}}[k(\kd{r}_{\kb{n}})]<0\ \forall \kd{r}_{\kb{n}}$ the minimum is located at $\kd{r}_{\kb{n,min}}=0$ with a depth of $\Delta\epsilon_{000}/\hbar \equiv [\varepsilon_{000}-\epsilon^{\kb{pt}}_{000}(\kd{r}_\kb{n,min})]/\hbar=591$ MHz. In the present case we have $A_{\kb{T}}[k(\kd{r}_{\kb{n}})] \approx A_{\kb{T}}(x^{2}_{\kb{n}}+y^{2}_{\kb{n}},z^{2}_{\kb{n}})$ and $a \approx b$, which means that the giant dipole ground state PES possesses an approximate $\varphi_{\kb{n}}$-rotational symmetry and $(L_{\kb{z}})_{\kb{n}}$ is an approximately conserved quantity. A harmonic fit $V_{\kb{h}}(\kd{r}_{\kb{n}})=\frac{1}{2}m_{\kb{n}}\omega^{2}_{x}x^{2}_{\kb{n}}+\frac{1}{2} m_{\kb{n}}\omega^{2}_{y}y^{2}_{\kb{n}}+\frac{1}{2}m_{\kb{n}}\omega^{2}_{z}z^{2}_{\kb{n}}$ around this minimum provides harmonic oscillator states with a level spacing of $\omega_{x}=\omega_{y}=39.2$ MHz and $\omega_{z}=12.4$ MHz. The numerical exact PES is presented in Fig.\ \ref{fig.1}(a). The employed basis set consists of $N_{-}=30,\ N_{+}=1,\ N_{\kb{z}}=50$ states. As in perturbation theory the potential possess a minimum located at $\kd{r}_{\kb{n,min}}=0$ and is $\varphi_{\kb{n}}$-rotationally symmetric. A harmonic fit provides here a depth of $\Delta\epsilon_{000}/\hbar=788$ MHz and level spacing of $\omega_{x}=\omega_{y}=20$ MHz, $\omega_{z}=6.5$ MHz.

In Fig.\ \ref{fig.1}(b) we show a comparison between the perturbative and the exact ground state potential for $y_{\kb{n}}=z_{\kb{n}}=0$. In addition, the harmonic fitting curves and the parameters $\omega^{\kb{pt}}_{x}$ and $\omega^{\kb{ex}}_{x}$ are presented as well. Comparing the analytic solution eq.\ (\ref{ptexc1}) with the numerically exact result, one finds that for the exact PES the depth possesses a $165$ MHz larger value while the width of the well increases by a factor of two. Furthermore, for the perturbative PES the harmonic approximation is valid in a region between $x_{\kb{n}}\approx\pm 200a_0$, while for the exact potential it is valid up to $x_{\kb{n}}\approx\pm 800a_0$.
Hence, while providing a good qualitative prediction, the perturbative approach cannot be used to discuss quantitative details. The deviation between the exact and the perturbative result can be understood by analyzing the coupling of the giant dipole groundstate to excited states of the unperturbed system. For first order perturbation theory to hold, these couplings need to be much smaller than the energetic separation of the involved levels. In our case, this amounts to the requirement
\begin{equation}
 \gamma_{n_{-}n_{+}n_{\kb{z}}}:=4 \left|\frac{\langle 000 | V_{\kb{n,GD}}(\kd{r},\kd{r}_{\kb{n}}) | n_{-}n_{+}n_{\kb{z}} \rangle_{\kd{r}}}{\epsilon^{\kb{pt}}_{000}(\kd{r}_{\kb{n}})-\epsilon^{\kb{pt}}_{n_{-}n_{+}n_{\kb{z}}}(\kd{r}_{\kb{n}})}\right|^2 \ll 1
\end{equation}
where $n_{-}$, $n_{+}$ and $n_{\kb{z}}$ label the quantum number of the excited state. In our case, we find, e.g., $\gamma_{100}\approx 4.5$ and $\gamma_{002}\approx0.1$. Hence, it is not surprising that first order perturbation theory yields qualitatively but not quantitatively reliable results. This is in contrast to the trilobite systems \cite{Greene00} where degenerate first order perturbation theory within a given Rydberg $n$-manifold provides satisfactory results (depending on $n$, the splitting of adjacent manifolds is in the GHz-THz regime compared to $\omega_{-}=223\ \text{MHz}$ in our case).


We remark that the vibrational states obtained by the potential given in Fig.\ 1 are localized at internuclear distances in the range of $10^{5}a_0$. To our knowledge, these molecules belong, together with recently investigated Rydberg macrodimers \cite{cote11}, to the largest diatomic molecules ever predicted.
\section{Potential surfaces of excited states}
\begin{figure}
\includegraphics[width=\columnwidth]{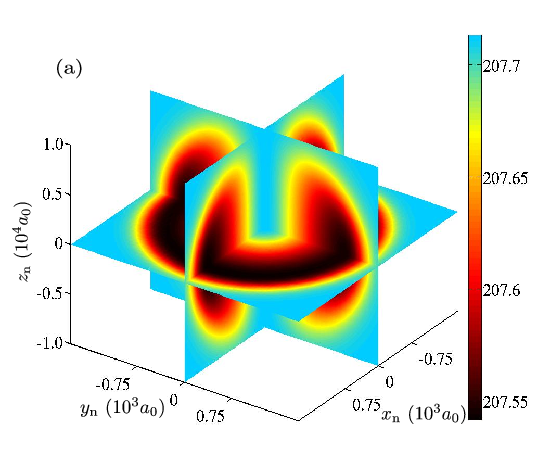}\\
\includegraphics[width=\columnwidth]{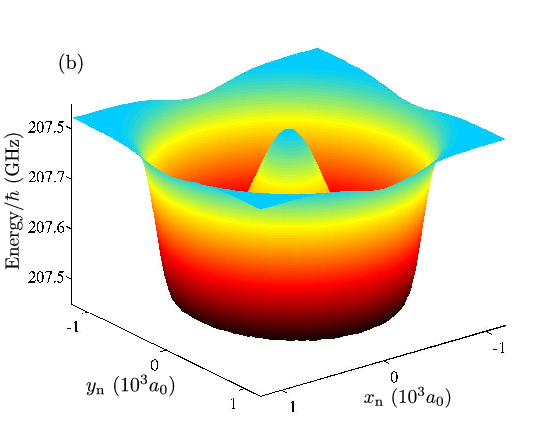}
\caption{First excited potential surface. To achieve convergence we use a set of  $N_{-}=10,\ N_{+}=1,\ N_{\kb{z}}=30$ basis functions.
(a) 3D representation and (b) 2D intersection for $z_{\kb{n}}=0$. The energy scale is given in GHz. (color online)}
\label{fig.3}
\end{figure}

In perturbation theory the PES for the lowest excitations are determined by the states $|n_{-} 00 \rangle,\ n_{-}=1,2,3$. Introducing cylindrical coordinates $\rho_{\kb{n}},\ \varphi_{\kb{n}},\ z_{\kb{n}}$ and extracting the dominant terms in perturbation theory, the PES are approximately given by 
\begin{eqnarray}
\epsilon^{\kb{pt}}_{n_{-}00}(\kd{r}_{\kb{n}}) \approx \varepsilon_{n_{-}00}+2\pi A_{\kb{T}}[k(\kd{r}_{\kb{n}})]C_{n_{-}}
e^{-a\rho^{2}_{\kb{n}}-cz^{2}_{\kb{n}}} \rho^{2n_{-}}_{\kb{n}}.
\end{eqnarray}
Due to the weak dependence of $A_{\kb{T}}[k(\kd{r}_{\kb{n}})]$ on $k$ in our case, this well represents a $\varphi_{\kb{n}}$-rotationally symmetric torus with minima at $\kd{r}^{(n_-)}_{\kb{n,min}}=\rho^{(n_-)}_{\kb{min}}\kd{e}_{\rho}$. The positions $\rho^{(n_-)}_{\kb{min}}$ and depths $\Delta\epsilon_{n_-00}$ of these minima are approximately given by 
\begin{equation}
\begin{split}
\rho^{(n_-)}_{\kb{min}} &\approx \sqrt{n_{-}/a},\\ 
\Delta\epsilon_{n_-00} &\approx 2\pi |A_{\kb{T}}[k(\kd{r}^{(n_-)}_{\kb{n,min}})]|\frac{C_{n_{-}}}{e^{n_-}}\left(\frac{n_{-}}{a}\right)^{n_-}.
\end{split}\label{radiusdepth}
\end{equation}
In Fig.\ \ref{fig.3}(a) the exact PES for the first excited state is shown. The toroidal structure as predicted by perturbation theory is clearly visible.  
Fig.\ \ref{fig.3} (b) shows a 2D intersection for $z_{\kb{n}}=0$ of the same PES: A ``Mexican hat`` like potential well is obtained with the one-dimensional manifold of the minimum lying on a circle. For sufficiently small displacements from the minima the PES can be described by a harmonic approximation, i.e., a decentered oscillator potential $V_{\kb{h}}(\kd{r}_{\kb{n}})=\frac{1}{2}m_{\kb{n}}\omega^{2}_{\rho}(\rho_{\kb{n}}-\rho_{\kb{min}})^{2}+\frac{1}{2}m_{\kb{n}}\omega^{2}_{z}z^{2}_{\kb{n}}$. One therefore arrives at a free rotational motion of the Rb-atom in the $\varphi_{\kb{n}}$-direction. In perturbation theory the frequencies $\omega_{\rho}$ and $\omega_{z}$ are given by
\begin{eqnarray}
\omega^{(n_-)}_{\rho} &\approx& 2\sqrt{\frac{2 \pi a|A[k(\kd{r}^{(n_-)}_{\kb{n,min}})]|C_{n_-}}{m_{\kb{Rb}}e^{n_-}}  \left( \frac{n_-}{a} \right)^{n_-}},\\
\omega^{(n_-)}_{z} &\approx& 2\sqrt{\frac{\pi c|A[k(\kd{r}^{(n_-)}_{\kb{n,min}})]|C_{n_-}}{m_{\kb{Rb}}e^{n_-}} \left( \frac{n_-}{a} \right)^{n_-} }.\label{omegaz}
\end{eqnarray}
The corresponding eigenfunctions are given by the product $A(\rho_{\kb{n}})e^{im\varphi_{\kb{n}}}\phi_{l}(z_{\kb{n}}),\ m \in \mathbb{Z}$ and $l \in \mathbb{N}_0$, where $\phi_{l}$ denotes the $l$-th eigenfunction of the harmonic oscillator and $A(\rho_{\kb{n}})$ are, apart from a Gaussian, the biconfluent Heun functions \cite{Heun}. In table \ref{table1} the parameters $\rho_{\kb{min}},\ \omega_{\rho}$ and $\omega_{\kb{z}}$ for the first three excited PES, both exact and perturbative, are listed. We see that the largest difference between perturbation theory and the numerically exact result is obtained for the $\rho_{\kb{min}}$ and $\omega_z$ parameters with a deviation of around $50 \%$ for the first excitation and $20 \%$ for the third excitation. However, for $\omega_{\rho}$ and $\Delta\epsilon_{n_-00}$ both results are more comparable with a maximum deviation of $20 \%$. For all quantities the exact and perturbative results are more comparable with increasing excitation. 

\begin{table}
\begin{center}
\begin{tabular}[c]{|c|c|c|c|c|c|c|}
\hline
× & \multicolumn{2}{|c|}{exc. 1} & \multicolumn{2}{|c|}{exc. 2} & \multicolumn{2}{|c|}{exc. 3}\\
\hline
× & pt & ex & pt  & ex & pt  & ex \\
\hline
$\rho_{\kb{min}}\ / \ a_0$ & 450 & 825 & 637 & 971 & 785 & 1050\\
\hline
$\Delta\epsilon_{n_-00}\ /\ \text{MHz}$ & 221 & 170 & 162 & 155 & 134 & 127\\
\hline
$\omega_\rho\ /\ \text{MHz}$ & 33 & 27 & 29 & 25 & 26 & 24\\
\hline
$\omega_z\ /\ \text{MHz}$ & 1.9 & 0.86 & 1.7 & 0.89 & 1.5 & 1.2\\
\hline
\end{tabular}
\caption{Parameters $\rho_{\kb{min}},\ \Delta\epsilon_{n_-00},\ \omega_{\rho}$ and $\omega_{\kb{z}}$ for the first three excited potential surfaces. For the numerically exact potentials (ex) the parameters are taken from a harmonic fitting. For the perturbative potentials (pt), the parameters are extracted from eqs.\ (\ref{radiusdepth})-(\ref{omegaz}).}
\label{table1}
\end{center}
\end{table}

\section{Avoided crossings of potential surfaces}
\begin{figure}
\includegraphics[width=\columnwidth]{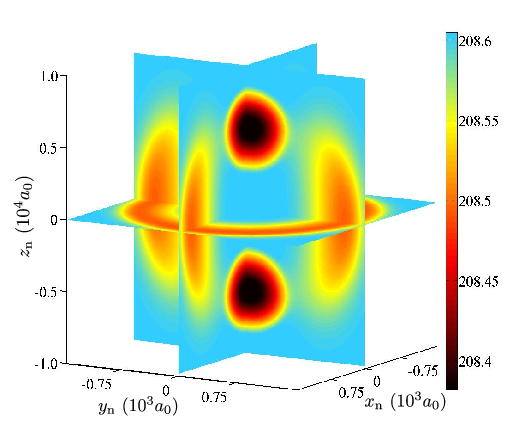}
\caption{Potential surface for the fifth excitation. The additional elliptical wells arise due to avoided crossings with adjacent potential surfaces. The energy scale is given in GHz. (color online)} 
\label{fig.5}
\end{figure}

While for the first few excitations first order perturbation theory provides a reasonable prediction of the qualitative behavior of the PES, for higher excitations first order perturbation theory is not capable of describing even the qualitative behavior of the PES. For example, in Fig.\ \ref{fig.5} we show the PES for the fifth excitation. In addition to the toroidal well described above we get two new 3D elliptical potential wells centered at $x_{\kb{n}}=y_{\kb{n}}=0,\ z_{\kb{n}}\approx\pm 6500 a_0$. As a consequence the former rotationally symmetric global minimum represent now a local minimum and the two new global minima are the centers of these elliptical wells. The bound states in the toroidal well become metastable and can in principle decay into bound vibrational states in the elliptical wells.

For the fifth and sixth excited PES a harmonic approximation around the global minima yields a vibrational level spacing of approximately $20$ MHz.   
\begin{figure}
\includegraphics[width=\columnwidth]{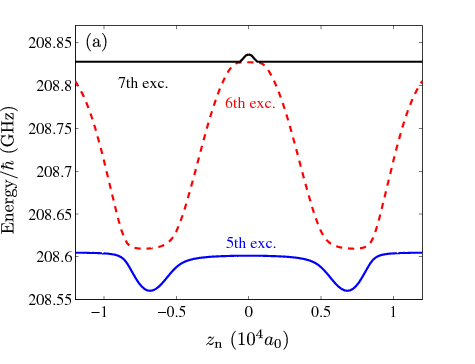}
\includegraphics[width=\columnwidth]{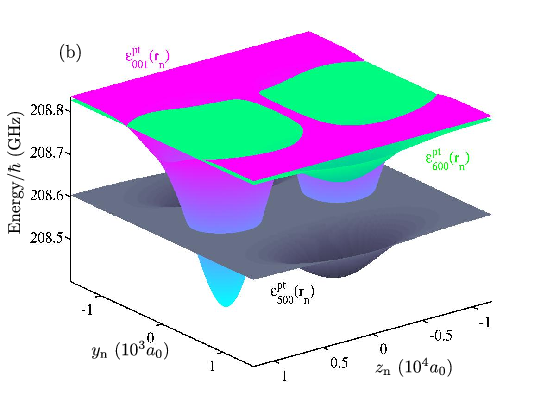}
\caption{(a) Intersections of the fifth, sixth and seventh excited potential surfaces for $x_{\kb{n}}=0,\ y_{\kb{n}}=200a_{0}$ are shown. Avoided crossings of the potentials are encountered. The spacing between the fifth and sixth surface at the avoided crossings is approximately 9 MHz. Subfigure (b) shows 2D intersections of the perturbatively calculated PES for $x_{\kb{n}}=0$. (color online)} 
\label{fig.6}
\end{figure}
The additional elliptical wells arise due to avoided crossing with neighboring PES. To confirm this, Fig.\ \ref{fig.6}(a) shows intersections for the fifth, sixth and seventh PES for $x_{\kb{n}}=0,\ y_{\kb{n}}=200a_{0}$. Avoided crossings of the PES are encountered, e.g.,\ at $z \approx \pm 8.5 \times 10^{3} a_0$ and $z \approx \pm 5.5 \times 10^{3} a_0 $ for the fifth and sixth excited PES.

A more global view of the geometry of the higher excited PES and their avoided crossings is provided with Fig.\ \ref{fig.6}(b). It shows the PES obtained via perturbation theory, i.e., $\epsilon^{\kb{pt}}_{500}(\kd{r}_{\kb{n}}),\ \epsilon^{\kb{pt}}_{600}(\kd{r}_{\kb{n}})$ and $\ \epsilon^{\kb{pt}}_{001}(\kd{r}_{\kb{n}})$ for fixed $x_{\kb{n}}=0$. The potential $\epsilon_{001}(\kd{r}_{\kb{n}})$ is obtained by raising the quantum number $n_{\kb{z}}$ and is given by
\begin{eqnarray}
\epsilon^{\kb{pt}}_{001}(\kd{r}_{\kb{n}})=\varepsilon_{001}+2\pi A_{\kb{T}}[k(\kd{r}_{\kb{n}})]\tilde{C}z^2e^{-a\rho^{2}_{\kb{n}}-cz^{2}_{\kb{n}}}.
\end{eqnarray}
In contrast to $\epsilon^{\kb{pt}}_{500}(\kd{r}_{\kb{n}})$ and $\epsilon^{\kb{pt}}_{600}(\kd{r}_{\kb{n}})$ this potential curve possess two minima at $z=\pm 5542a_0$. Because the depth of these minima ($\approx 450$ MHz) is larger than the level spacing of the unperturbed giant dipole levels [$(\varepsilon_{001}-\varepsilon_{600})/\hbar \approx 9\ \text{MHz},\ (\varepsilon_{001}-\varepsilon_{500})/\hbar \approx 250$ MHz] the PES $\epsilon^{\kb{pt}}_{001}(\kd{r}_{\kb{n}})$ intersects neighboring PES. Due to the coupling between the different PES this leads to avoided crossings for the exact PES and consequently novel geometries of the potentials. In the vicinity of the avoided crossing the adiabatic approximation fails and we expect a strong rovibronic interaction mixing different electronic giant dipole states. As a consequence fast decay processes of wave packets probing the seam of the avoided crossings will take place. Finally we note that the radiative lifetimes of the excited states are for our chosen parameter values of the order of several days and therefore much longer than the typical vibrational frequencies in the excited PES: it should therefore be possible to probe the vibrational dynamics belonging to the complex geometry of the molecular PES of the excited giant dipole states.

We remark that the occurence of avoided crossing will be even more prominent in heavier giant dipole systems. Considering rubidium instead of hydrogen, for example, reduces the smallest level spacing by $m_{p}/M_{\kb{Rb}} \approx 10^{-2}$, which yields $\omega_{-}=2.5$ MHz. The depths of the potential surfaces, on the other hand, do not change significantly. As a result, already the ground state shows avoided crossings with neighboring potential surfaces.


\section{Conclusion}
Bringing together the concepts underlying atomic giant dipole states in external fields and ultra-long-range molecules, we demonstrated the existence of ultra-long-range giant dipole diatomic molecules. In particular, the exotic atomic state underlying these molecules give rise to novel properties such as a plethora of different quantum states with complex three-dimensional energy landscapes and rich rovibrational dynamics. The resulting molecules possess very large rovibrational bound states at internuclear distances in the range of $10^5a_0$.

For their experimental preparation the 'best of two worlds' has to
be combined. The preparation of giant dipole states is known to be possible starting from 'traditional' Rydberg
states in magnetic fields and applying a sequence of electric field switches which brings the electron into low-lying
outer well states \cite{Averbukh99}. Driven radio-frequency transitions in the outer well as an additional tool might help to prepare 
definite outer well states. Starting from these, one could overlap the GDS with a dense cloud of ultracold rubidium atoms and use radio- / microwave induced transitions to form the envisaged giant dipole molecular states. One of the main differences to standard cold atom experiments is certainly the regime of field strengths necessary to address the giant dipole states, which corresponds to strong static magnetic and electric fields.
\section{Acknowledgment}
M.M. acknowledges financial support by a fellowship within the postdoc-programme of the German Academic Exchange Service (DAAD).
\end{document}